\documentclass{article}
\usepackage{spconf,amsmath,graphicx}
\usepackage{hyperref}       
\usepackage{url}            
\usepackage{booktabs}       
\usepackage{amsfonts}       
\usepackage{nicefrac}       
\usepackage{microtype}      
\usepackage{xcolor}
\usepackage{amsmath,amsthm,amssymb}
\usepackage{graphicx}
\usepackage{svg,tikz}
\usepackage{amsthm}
\usepackage{enumitem}

\usepackage{numprint}
\usepackage{siunitx}
\usepackage{tikz}
\usepackage{pgfmath}
\usepackage{pdfpages} 

\usepackage[normalem]{ulem}

\usepackage[utf8]{inputenc} 
\usepackage[T1]{fontenc}    

\title{A Gated Hypernet Decoder for Polar Codes}
\name{Eliya Nachmani, Lior Wolf}
\address{Facebook AI Research and Tel-Aviv University}
\begin{document}
%
\maketitle
\begin{abstract}
Hypernetworks were recently shown to improve the performance of message passing algorithms for decoding error correcting codes. In this work, we demonstrate how hypernetworks can be applied to decode polar codes by employing a new formalization of the polar belief propagation decoding scheme. We demonstrate that our method improves the previous results of neural polar decoders and achieves, for large SNRs, the same bit-error-rate performances as the successive list cancellation method, which is known to be better than any belief propagation decoders and very close to the maximum likelihood decoder.
\end{abstract}
\section{Introduction}
\label{sec:intro}

The development of neural decoders for error correcting codes has been evolving along multiple axes. In one axis, learnable parameters have been introduced to increasingly sophisticated decoding methods. Polar codes, for example, benefit from structural properties that require more dedicated message passing methods than conventional LDPC decoders. A second axis is that of the role of learnable parameters. Initially, weights were introduced to existing computations. Subsequently, neural networks replaced some of the computations and generalized these. The introduction of hypernetworks, in which the weights of the network vary based on the input, added a new layer of adaptivity. 

In this work, we address the specialized belief propagation decoder for polar codes of Arikan~\cite{arikan2010polar}, which makes use of the structural properties of these codes. We generalize the work of Xu et al.~\cite{xu2017improved} and Teng et al.~\cite{teng2019low}, both built upon the same underlying decoder, by introducing a graph neural network decoder whose architecture varies, as well as it weights. This allows our decoder to better adapt to the input signal.

We demonstrate our results on polar codes of various block sizes and show improvement in all SNRs over the baseline methods. Furthermore, for large SNRs, our method matches the performance of the successive list cancellation decoder~\cite{tal2011list}.

\section{Related Work}
\label{sec:related_work}

\noindent{\bf Network decoders\quad}
Deep learning techniques are increasingly applied to decode error correcting codes. Vanilla fully connected neural networks were applied for polar code decoding~\cite{gruber2017deep}. The performance obtained for short polar codes, e.g., $n=16$ bits, was close to that of the optimal maximum a posteriori (MAP) decoding. The fully connected networks, however, do not scale well, since the number of codewords grows exponentially in the number of information bits $k$.

For decoding larger block codes ($n\geqslant100$), message passing methods were augmented with learned parameters. For example, in~\cite{nachmani2016learning} the belief propagation (BP) decoding method is reincarnated as a neural network with weights that are assigned to each variable edge. A hardware-friendly BP method was introduced by employing the min-sum iterations~\cite{lugosch2017neural}. In both cases, an improvement is shown in comparison to the baseline BP method.

Vasic et al. learn the node activations based on components from existing decoders (BF, GallagerB, MSA, SPA)~\cite{vasic2018learning}. In \cite{kim2018communication}, a Recurrent Neural Network (RNN) decoder was shown to achieve close to optimal accuracy in decoding convolutional and Turbo codes, similar to the classical convolutional codes decoders Viterbi and BCJRI~\cite{kim2018communication}. 

Specifically for Polar codes, where the encoding graph can be partitioned into sub-blocks, Cammerer et al.~\cite{cammerer2017scaling} performed neural decoding  for each sub-block separately and then combine the sub-blocks and use a belief propagation decoding algorithm. Doan et al.~\cite{doan2018neural} improve this method by using successive cancellation decoding between the sub-blocks, and show a $42.5\%$ improvement in the decoding latency. Doan et al. further introduce in \cite{doan2019neural1} a neural decoder for the CRC-Polar code. The neural decoder uses trainable weights on the concatenated factor graph and shows an improvement of $0.5dB$ over the CRC-aided BP decoder. 
Xu et al.~\cite{xu2017improved} introduce weights on the edges on the polar belief propagation decoder of~\cite{arikan2010polar}. The training is done on the noisy zero codeword as in~\cite{nachmani2016learning}. Teng et al.~\cite{teng2019low} further improve the neural polar decoder and introduce an RNN polar decoder with a lower number of learnable parameters. Dai et al.~\cite{dai2018new} suggest a neural offset min-sum decoder for polar codes, and show that their decoder is more suitable for hardware implementation, since it uses addition instead of multiplication. Wodiany et al. improve the method of Dai et al., by presenting a new low-precision neural decoder and show that their method reduces by a factor of 54 the number of weights~\cite{wodiany2019low}. Xu et al.~\cite{xu2018polar1} employ a single weight for a min-sum neural polar decoder, which is used as a unified weight for all edges that connect check nodes to variable nodes. Their method shows comparative results to the sum-product algorithm. 

\noindent{\bf hypernetworks\quad} Dynamic networks, also know as hypernetworks, are schemes in which one network $f$ predicts the weights $\theta_h$ of another network {$h$}. This was first used for specific layers in order to adapt the network to motion or illumination changes~\cite{klein2015dynamic,7410424}. More layers were subsequently employed to interpolate video frames and predict stereo views~\cite{jia2016dynamic}. Hypernetworks in which both $f$ and $h$ are RNNs, were used for state of the art sequence modeling~\cite{ha2016hypernetworks}. Hypernetworks can also be naturally applied to metalearning, since $f$ can share information between tasks~\cite{bertinetto2016learning}. Another advantage is the ability to generate network weights on the fly without backpropagation, which is useful in neural architecture search (NAS)~\cite{brock2018smash,zhang2018graph}. 

In a recent application of hypernetworks to graphs, we used an MLP to generate the weights of a message passing network that decodes LDPC, BCH, and Polar error correcting codes~\cite{nachmani2019}. This generalizes earlier attempts in the domain of network decoders mentioned above and is shown to greatly improve performance over~\cite{nachmani2017learning}. The input to both the weight generating network $f$ and the message generation network $h$  is the incoming message, where for the first network, the absolute value is used, in order to  focus on the confidence of each message, rather than on the content of the message and maintaining the symmetry conditions required for efficient training. Hypernetworks often suffer from severe initialization challenges, since the weights of $h$ are determined by $f$ and do not follow a conventional initialization scheme. In that work, we, therefore, present a new activation function that is more stable than the $arctanh$ activation typically used in message passing decoders. 
In a followup work, we employ conventional activations, but suggest combining the initial message (from the first iteration) with the last message as an effective way to stabilize the training of the graph hypernetwork~\cite{nachmani2020molecule}. Performance is similar or better than what is obtained in~\cite{nachmani2019}.

In this work, we use the hypernetwork graph neural network scheme to augment the polar neural decoder from \cite{xu2017improved}. Furthermore, we modify the scheme to adapt the architecture of the graph neural network via a gating mechanism that is conditioned on the input. 

\vspace{-.2cm}
\section{Method}
\label{sec:method}
We begin by describing the evolution of  neural polar decoders and then describe our own method.
\subsection{Background}
Polar codes and their belief propagation decoder were first introduced by Arikan in \cite{arikan2008channel} and \cite{arikan2010polar} respectively. More precisely, assume that we have a $(N,K)$ polar code, where $N$ is the block size and $K$ is the number of information bits. The polar factor graph has $(n+1)N$ nodes, $\frac{N}{2}log_{2}N$ blocks and $n=log_{2}N$ stages. Each node in the factor graph index by tuple $(i, j)$ where $1 \leq i \leq n+1, 1 \leq j \leq N$. The rightmost nodes $(n+1,\cdot)$ are the noisy input from the channel $y_j$, and the leftmost nodes $(1,\cdot)$ are the source data bits $u_j$. The polar belief propagation decoder uses two types of messages in order to estimate the log
likelihood ratios (LLRs): left and right messages $L^{(t)}_{i,j}$, $R^{(t)}_{i,j}$, where $t$ is the number of the BP iteration. The left messages are initialized at $t=0$ with the input log likelihood ratio: 
\begin{equation}
\label{eq:init_l}
    L^{(1)}_{n+1,j}=ln\left (  \frac{P\left ( y_j  \mid x_j=0 \right )}{P\left ( y_j \mid x_j=1 \right )}\right )
\end{equation}
The right messages are initialized with the information bit location: 
\begin{equation}
\label{eq:init_r}
    R^{(1)}_{n+1,j}=\frac{P\left ( u_j=0 \right )}{P\left ( u_j=1 \right )}=\left\{\begin{matrix}
0 &  \textup{j is an information bit} \\ 
\infty  &  \textup{else}
\end{matrix}\right.
\end{equation}
The other messages $L^{(1)}_{i,j}$, $R^{(1)}_{i,j}$ are set to 1. The iterative belief propagation equation for the messages are:
\begin{equation}
\label{eq:polar_bp}
\begin{split}
&L^{(t)}_{i,j} = g\left ( L^{(t-1)}_{i+1,j}, L^{(t-1)}_{i+1,j+N_i} +  R^{(t)}_{i,j+N_i}  \right ) , \\ 
&L^{(t)}_{i,j+N_i} = g\left ( R^{(t)}_{i,j}, L^{(t-1)}_{i+1,j} \right ) + L^{(t-1)}_{i+1,j+N_i} , \\ 
&R^{(t)}_{i+1,j} =   g\left ( R^{(t)}_{i,j}, L^{(t-1)}_{i+1,j+N_i} + R^{(t)}_{i,j+N_i} \right ) ,  \\ 
&R^{(t)}_{i+1,j+N_i} = g\left ( R^{(t)}_{i,j}, L^{(t-1)}_{i+1,j} \right ) + R^{(t)}_{i,j+N_i}
\end{split}
\end{equation}where $N_i=N/2^{i}$ and the function $g$ is:
\begin{equation}
\label{eq:g_func}
g(x,y)=\textup{ln}\frac{1+xy}{x+y}.
\end{equation}
Alternatively, $g$ can be replaced by the min-sum approximation~\cite{pamuk2011fpga}:
\begin{equation}
\label{eq:g_func_app}
g(x,y)\approx \textup{sign(x)}\cdot \textup{sign(y)}\cdot\min(\left | x \right |, \left | y \right |)
\end{equation}
The final estimation is a hard slicer on the left messages $L^{(T)}_{1,j}$ where $T$ is the last iteration:
\begin{equation}
\label{eq:final_bp}
\hat{u}^{N}_{j}=\left\{\begin{matrix}
0, L^{(T)}_{1,j}\geqslant 0, \\  
1 , L^{(T)}_{1,j} < 0
\end{matrix}\right.
\end{equation}

Xu at el.~\cite{xu2017improved} introduce a neural polar decoder that unfolds the polar factor graph and assigns weights in each edge. The update equation is taking the form:
\begin{equation}
\label{eq:polar_bp_deep}
\begin{split}
&L^{(t)}_{i,j} = \alpha^{(t)}_{i,j} \cdot g\left ( L^{(t-1)}_{i+1,j}, L^{(t-1)}_{i+1,j+N_i} +  R^{(t)}_{i,j+N_i}  \right ) , \\ 
&L^{(t)}_{i,j+N_i} = \alpha^{(t)}_{i,j+N_i} \cdot g\left ( R^{(t)}_{i,j}, L^{(t-1)}_{i+1,j} \right ) + L^{(t-1)}_{i+1,j+N_i} , \\ 
&R^{(t)}_{i+1,j} =   \beta^{(t)}_{i+1,j} \cdot g\left ( R^{(t)}_{i,j}, L^{(t-1)}_{i+1,j+N_i} + R^{(t)}_{i,j+N_i} \right ) ,  \\ 
&R^{(t)}_{i+1,j+N_i} = \beta^{(t)}_{i+1,j+N_i} \cdot g\left ( R^{(t)}_{i,j}, L^{(t-1)}_{i+1,j} \right ) + R^{(t)}_{i,j+N_i}
\end{split}
\end{equation}

where $\alpha^{(t)}_{i,j}$ and $\beta^{(t)}_{i,j}$ are learnable parameters for the left message $L^{(t)}_{i,j}$ and right message $R^{(t)}_{i,j}$ respectively. The output of the neural decoder is defined by:
\begin{equation}
\label{eq:polar_bp_output}
o_j=\sigma\left ( L^{(T)}_{1,j} \right )
\end{equation}
where $\sigma$ is the sigmoid activation. The loss function was the cross entropy between the transmitted codeword and the network output:
\begin{equation}
\label{eq:polar_bp_loss}
L\left ( o,u \right )=-\frac{1}{N}\sum_{j=1}^{N}u_j\textup{log}(o_j)+(1-u_j)\textup{log}(1-o_j)
\end{equation}

Teng et al.~\cite{teng2019low} further introduced a recurrent neural polar decoder, by sharing the weights among different iterations:
$\alpha^{(t)}_{i,j}$ = $\alpha_{i,j}$ and $\beta^{(t)}_{i,j}$ = $\beta_{i,j}$. Their BER-SNR curve achieves comparable results to training the neural decoder without tying the weights from different iterations.
\subsection{The Hypernetwork Method}
In this work, we propose a new structure-adaptive hypernetwork architecture for decoding polar codes. The proposed architecture adds three major modifications. First, we incorporate a graph neural network that uses the unique structure of the polar code, following the work of \cite{nachmani2019}. Second, we suggest adding a gating mechanism to the activations of the (hyper) graph network, in order to adapt the architecture itself according to the input. Third, we add a damping factor $c$ to the updating equations following \cite{nachmani2020molecule} in order to improve the training stability of the proposed method. 

\begin{figure}[t]
\centering
\includegraphics[width=.49495\textwidth]{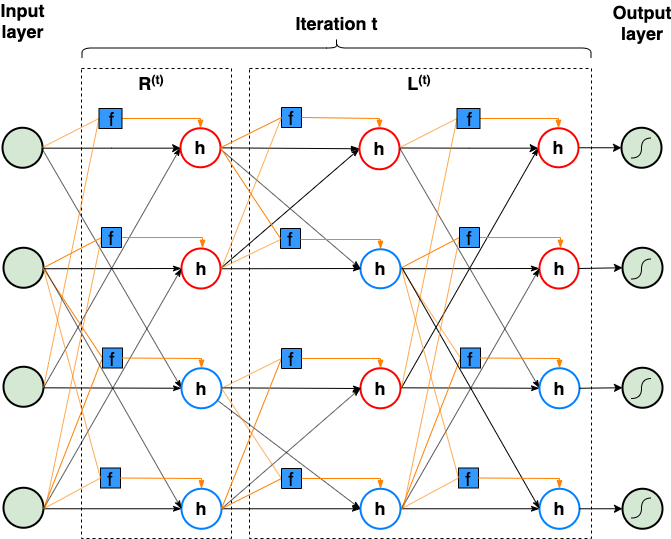}
\caption{An overview of our method for polar code with $N=4$ and $T=1$. The connections of the graph hypernetwork are denoted by orange lines. $f$ is the function that determines the weights of the graph nodes $h$. 
To reduce clutter, the damping factors are not shown.}
\label{fig:graph_hy}
\end{figure}

The method is illustrated in Fig.~\ref{fig:graph_hy}. At each iteration $t$, we employ the hyper-network $f$:
\begin{equation}
\label{eq:polar_ada_hyper_theta}
\begin{split}
&\theta^{(t)}_{i,j}, \sigma^{(t)}_{i,j}  = f\left ( |L^{(t-1)}_{i+1,j}|, |L^{(t-1)}_{i+1,j+N_i} +  R^{(t)}_{i,j+N_i}| \right )\\
&\theta^{(t)}_{i,j+N_i}, \sigma^{(t)}_{i,j+N_i}  = f\left ( |R^{(t)}_{i,j}|, |L^{(t-1)}_{i+1,j}| \right )\\
&\theta^{(t)}_{i+1,j}, \sigma^{(t)}_{i+1,j}  = f\left (  |R^{(t)}_{i,j}|, |L^{(t-1)}_{i+1,j+N_i} + R^{(t)}_{i,j+N_i}|  \right )\\
&\theta^{(t)}_{i+1,j+N_i}, \sigma^{(t)}_{i+1,j+N_i}  = f\left (  |R^{(t)}_{i,j}|, |L^{(t-1)}_{i+1,j}|  \right )\\
\end{split}
\end{equation}

where $f$ is a neural network that determines the weights and gating activation of network $h$. The network $f$ has four layers with $tanh$ activations. Note that the inputs to the function $f$ are in absolute value. We use the absolute value of the input messages in order to focus on the correctness of the messages and not the bit information, see \cite{nachmani2019} for more details.  

Furthermore, we replace the updating Eq.~\ref{eq:polar_bp_deep} with the following equations:
\begin{equation}
\label{eq:polar_ada_hyper_g}
\begin{split}
&L^{(t)}_{i,j} = (1-c)\cdot h\left ( L^{(t-1)}_{i+1,j}, L^{(t-1)}_{i+1,j+N_i} +  R^{(t)}_{i,j+N_i}, \theta^{(t)}_{i,j} , \sigma^{(t)}_{i,j} \right )\\ &+c \cdot \alpha^{(t)}_{i,j} \cdot g\left ( L^{(t-1)}_{i+1,j}, L^{(t-1)}_{i+1,j+N_i} +  R^{(t)}_{i,j+N_i}  \right ) , \\ 
&L^{(t)}_{i,j+N_i} = (1-c)\cdot h\left ( R^{(t)}_{i,j}, L^{(t-1)}_{i+1,j}, \theta^{(t)}_{i,j+N_i}, \sigma^{(t)}_{i,j+N_i} \right )\\ &+c \cdot \alpha^{(t)}_{i,j+N_i} \cdot g\left ( R^{(t)}_{i,j}, L^{(t-1)}_{i+1,j} \right ) + L^{(t-1)}_{i+1,j+N_i} , \\ 
&R^{(t)}_{i+1,j} = (1-c)\cdot h\left ( R^{(t)}_{i,j}, L^{(t-1)}_{i+1,j+N_i} + R^{(t)}_{i,j+N_i}, \theta^{(t)}_{i+1,j}, \sigma^{(t)}_{i+1,j} \right )\\ &+c \cdot \beta^{(t)}_{i+1,j} \cdot g\left ( R^{(t)}_{i,j}, L^{(t-1)}_{i+1,j+N_i} + R^{(t)}_{i,j+N_i} \right ) ,  \\ 
&R^{(t)}_{i+1,j+N_i} = (1-c)\cdot h\left ( R^{(t)}_{i,j}, L^{(t-1)}_{i+1,j}, \theta^{(t)}_{i+1,j+N_i}, \sigma^{(t)}_{i+1,j+N_i} \right )\\ &+c \cdot \beta^{(t)}_{i+1,j+N_i} \cdot g\left ( R^{(t)}_{i,j}, L^{(t-1)}_{i+1,j} \right ) + R^{(t)}_{i,j+N_i}
\end{split}
\raisetag{15pt}
\end{equation}
where the damping factor $c$ is a learnable parameter, initialized from uniform distribution $[0,1]$ and learned with clipping to the range of $[0,1]$ during the training. The network $h$ has two layers  with $tanh$ activations. 
Note that the weights of network $h$ are determined by the network $f$, and the activations of each layer in $h$ are multiplied by the gating $\sigma^{(t)}_{i,j}$ from the network $f$. The output layer and the loss function is the same as in Eq.~\ref{eq:polar_bp_output} and Eq.~\ref{eq:polar_bp_loss} respectively.

It is straightforward to show that the conditions of Lemma 2 of~\cite{nachmani2019} hold in our case as well and, therefore, the decoding error is independent of the transmitted codeword, allowing training solely with noisy versions of the zero codeword.

\section{Experiments}
\label{sec:experiments}
In order to evaluate our method, we train the proposed neural network for Polar codes with different block sizes $N=128,32$. The number of iterations was $T=5$ for all block codes.  The $f$ and $h$ networks have $16$ neurons in each layer, with $tanh$ activations and without a bias term. 
We generate the training set of noisy variations of the zero codeword over an additive white Gaussian noise channel (AWGN).  Each batch contains multiple examples from different Signal-To-Noise (SNR) values, specifically we use SNRs values of $1dB,2dB,..,6dB$. 
A batch size of $3600$ and $1800$ examples is used for $N=32$ and $N=128$, respectively. Learning rate at epoch $k$ is set according to $lr_{k} = lr_{0}/(1+k \cdot decay)$ where $lr_{0}=0.99$ and $lr_{0}=2.5$ for $N=32$ and $N=128$ respectively. The decay factor was $1e-4$ and every epoch contain $125$ batches. In all experiments, we use the feed-forward neural decoder, and do not use weight tying as in \cite{teng2019low}.

Similar to previous work~\cite{teng2019low}, the BER calculation uses the information bits, i.e. we do not count the frozen bits when calculating the error rate performance.

We compare our method with the vanilla belief propagation algorithm, the neural polar decoder of Xu et al. \cite{xu2017improved} and the successive list cancellation (SLC) method \cite{tal2011list} which does not employ learning and obtains state of the art performance.

\begin{figure}[t]
\centering
\includegraphics[width=.4569495\textwidth]{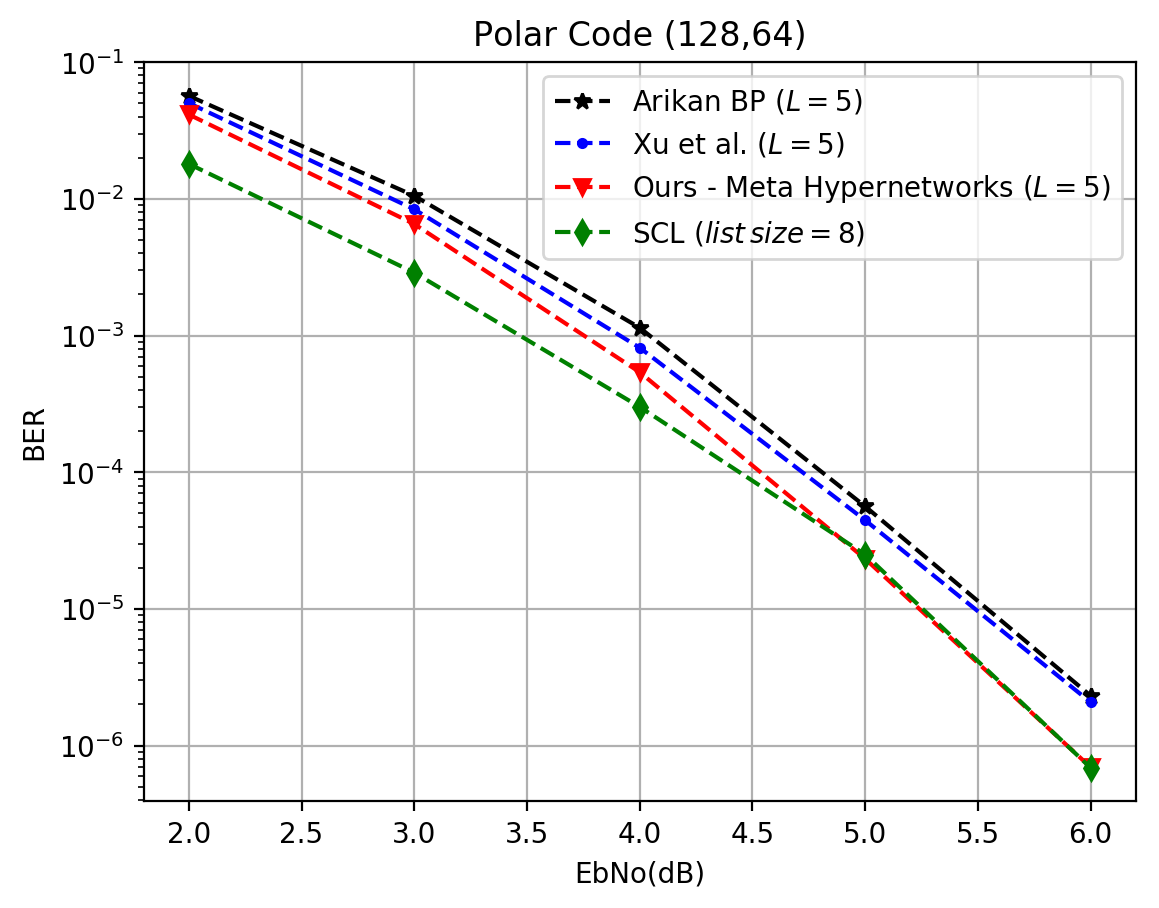}
\vspace{-.4352cm}
\caption{BER for Polar code (128,64)}
\label{fig:polar_128}
\end{figure}

\begin{figure}[t]
\centering
\includegraphics[width=.4569495\textwidth]{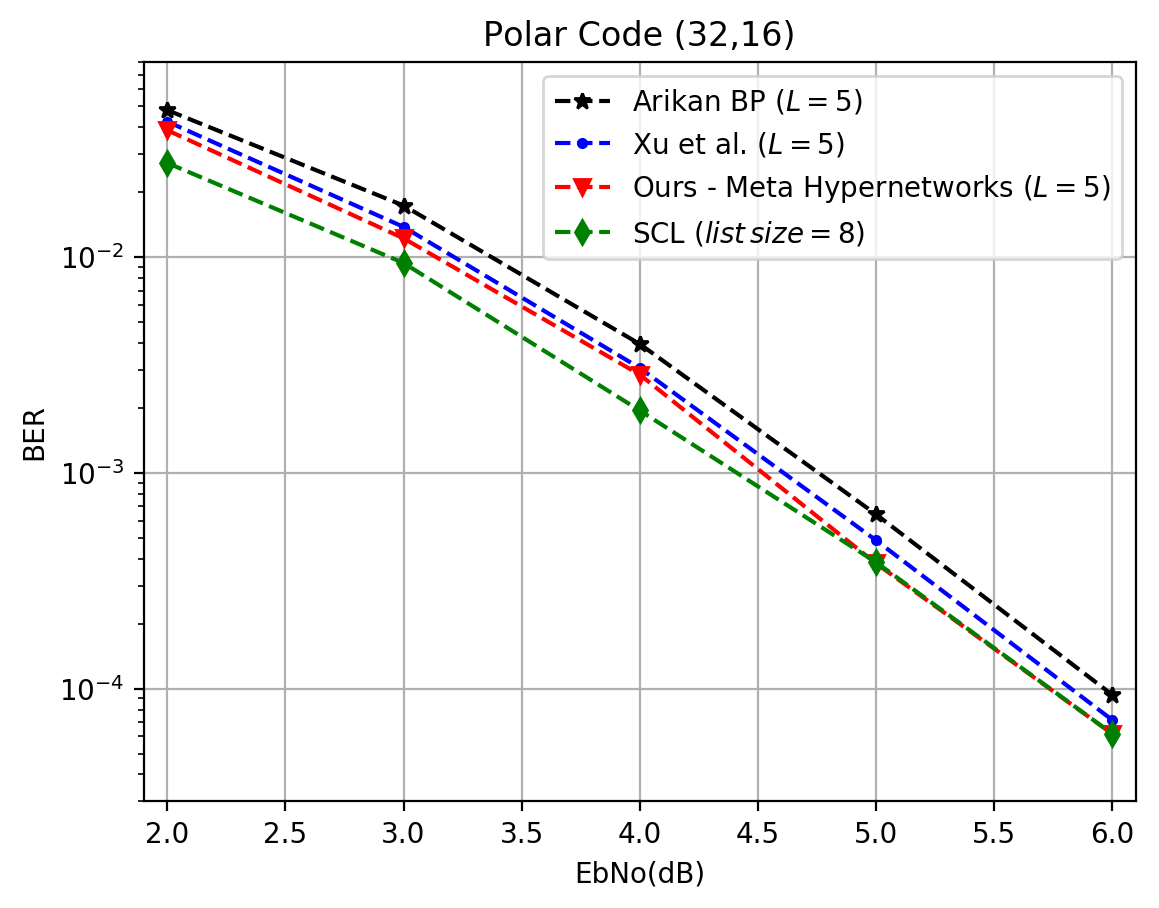}
\vspace{-.4352cm}
\caption{BER for Polar code (32,16)}
\label{fig:polar_32}
\end{figure}

In Fig.~\ref{fig:polar_128} and Fig.~\ref{fig:polar_32} we present the Bit-Error-Rate versus EbN0 for $N=128$ and $N=32$, respectively. As can be seen, for $N=32$ our method's accuracy matches that of SLC for large SNRs ($5dB,6dB$). Furthermore, for lower SNRs, our method improves the results of Xu et al. \cite{xu2017improved} by $0.1dB$. For large block, $N=128$, one can observe the same improvement in large SNRs value, where our method achieves the same performance as SLC, which is $0.4dB$ better than the baseline method of Xu et al. For lower SNRs, our method improves Xu et al. by $0.2dB$.

In order to evaluate the contribution of the various components of our method, we run an ablation analysis: (i) without the damping factor (ii) when using a fixed $c=0.5$ in Eq.~\ref{eq:polar_ada_hyper_g} (iii) without the gating mechanism (iv) the complete method. We run the ablation study on a polar code with $N=32$. 

Tab.\ref{tab:ablation} reports the results of the ablation analysis. As can be observed, the complete method, including the gating mechanism, outperforms a similar method without the damping factor (i) and without the gating mechanism (iii). Moreover, for training without the damping factor, the performance is equal to a random guess. Training with a fixed $c=0.5$  damping factor (ii) produces better results than $c=0$, however these results are worse then the complete method (iv).
\begin{table}[t]
    \centering
    \caption{Ablation analysis for polar code $(32,16)$. The negative natural logarithm of BER results of our complete method are compared with several variants. Higher is better.}
    \label{tab:ablation}
    \begin{tabular}{@{}lc@{~}c@{~}c@{~}c@{~}c@{}}
    \toprule
        Variant/SNR$[dB]$ & 1 & 2 & 3 & 4 & 5\\ 
        \midrule  
        (i) No damping factor $c=0$ & 0.73 & 0.73 & 0.74 & 0.74 & 0.75 \\
        (ii) Unlearned damping $c=0.5$ & 1.19 & 1.52 & 2.00 & 2.65 & 3.47 \\
        (iii) No gating mechanism & 2.39 & 3.20 & 4.36 & 5.81 & 7.75 \\
        (iv) Complete method& 2.42 & 3.25 & 4.40 & 5.85 & 7.87 \\
        \bottomrule 
    \end{tabular}
\end{table}

\section{Conclusions}
\label{sec:conclusions}
We present a hypernetwork scheme for decoding polar codes with a graph neural network. A novel gating mechanism is added in order to allow the network to further adapt to the input. We demonstrate our results on various polar codes and show that our method can achieve the same performance as successive list cancellation for large SNRs. 
\subsection*{Acknowledgments}
We thank Chieh-Fang Teng for sharing code and helpful discussions. The contribution of Eliya Nachmani is part of a Ph.D. thesis at Tel Aviv University.

\vfill\pagebreak

\bibliographystyle{IEEEbib}
\bibliography{refs}

\begin{thebibliography}{10}

\bibitem{arikan2010polar}
Erdal Arikan,
\newblock ``Polar codes: A pipelined implementation,''
\newblock in {\em ISBC}, 2010, pp. 11--14.

\bibitem{xu2017improved}
Weihong Xu, Zhizhen Wu, Yeong-Luh Ueng, Xiaohu You, and Chuan Zhang,
\newblock ``Improved polar decoder based on deep learning,''
\newblock in {\em SiPS}. IEEE, 2017, pp. 1--6.

\bibitem{teng2019low}
Chieh-Fang Teng, Chen-Hsi~Derek Wu, Andrew Kuan-Shiuan Ho, and An-Yeu~Andy Wu,
\newblock ``Low-complexity recurrent neural network-based polar decoder with
  weight quantization mechanism,''
\newblock in {\em ICASSP}. IEEE, 2019, pp. 1413--1417.

\bibitem{tal2011list}
Ido Tal and Alexander Vardy,
\newblock ``List decoding of polar codes,''
\newblock in {\em 2011 ISIT}. IEEE, 2011, pp. 1--5.

\bibitem{gruber2017deep}
Tobias Gruber, Sebastian Cammerer, Jakob Hoydis, and Stephan ten Brink,
\newblock ``On deep learning-based channel decoding,''
\newblock in {\em CISS}. IEEE, 2017, pp. 1--6.

\bibitem{nachmani2016learning}
Eliya Nachmani, Yair Be'ery, and David Burshtein,
\newblock ``Learning to decode linear codes using deep learning,''
\newblock in {\em 2016 54th Annual Allerton Conference on Communication,
  Control, and Computing (Allerton)}. IEEE, 2016, pp. 341--346.

\bibitem{lugosch2017neural}
Loren Lugosch and Warren~J Gross,
\newblock ``Neural offset min-sum decoding,''
\newblock in {\em ISIT}. IEEE, 2017, pp. 1361--1365.

\bibitem{vasic2018learning}
Bane Vasi{\'c}, Xin Xiao, and Shu Lin,
\newblock ``Learning to decode ldpc codes with finite-alphabet message
  passing,''
\newblock in {\em ITA}. IEEE, 2018, pp. 1--9.

\bibitem{kim2018communication}
Hyeji Kim, Yihan Jiang, Ranvir Rana, Sreeram Kannan, Sewoong Oh, and Pramod
  Viswanath,
\newblock ``Communication algorithms via deep learning,''
\newblock in {\em ICLR}, 2018.

\bibitem{cammerer2017scaling}
Sebastian Cammerer, Tobias Gruber, Jakob Hoydis, and Stephan ten Brink,
\newblock ``Scaling deep learning-based decoding of polar codes via
  partitioning,''
\newblock in {\em GLOBECOM}. IEEE, 2017, pp. 1--6.

\bibitem{doan2018neural}
Nghia Doan, Seyyed~Ali Hashemi, and Warren~J Gross,
\newblock ``Neural successive cancellation decoding of polar codes,''
\newblock in {\em SPAWC}. IEEE, 2018, pp. 1--5.

\bibitem{doan2019neural1}
Nghia Doan, Seyyed~Ali Hashemi, Elie~Ngomseu Mambou, Thibaud Tonnellier, and
  Warren~J Gross,
\newblock ``Neural belief propagation decoding of crc-polar concatenated
  codes,''
\newblock in {\em ICC}. IEEE, 2019, pp. 1--6.

\bibitem{dai2018new}
Bin Dai, Rongke Liu, and Zhiyuan Yan,
\newblock ``New min-sum decoders based on deep learning for polar codes,''
\newblock in {\em SiPS}. IEEE, 2018, pp. 252--257.

\bibitem{wodiany2019low}
Igor Wodiany and Antoniu Pop,
\newblock ``Low-precision neural network decoding of polar codes,''
\newblock in {\em SPAWC}. IEEE, 2019, pp. 1--5.

\bibitem{xu2018polar1}
Weihong Xu, Xiaohu You, Chuan Zhang, and Yair Be’ery,
\newblock ``Polar decoding on sparse graphs with deep learning,''
\newblock in {\em ACSSC}. IEEE, 2018, pp. 599--603.

\bibitem{klein2015dynamic}
Benjamin Klein, Lior Wolf, and Yehuda Afek,
\newblock ``A dynamic convolutional layer for short range weather prediction,''
\newblock in {\em CVPR}, 2015, pp. 4840--4848.

\bibitem{7410424}
G.~{Riegler}, S.~{Schulter}, M.~{Rüther}, and H.~{Bischof},
\newblock ``Conditioned regression models for non-blind single image
  super-resolution,''
\newblock in {\em ICCV}, Dec 2015, pp. 522--530.

\bibitem{jia2016dynamic}
Xu~Jia, Bert De~Brabandere, Tinne Tuytelaars, and Luc~V Gool,
\newblock ``Dynamic filter networks,''
\newblock in {\em NIPS}, 2016.

\bibitem{ha2016hypernetworks}
David Ha, Andrew Dai, and Quoc~V Le,
\newblock ``Hypernetworks,''
\newblock {\em arXiv preprint arXiv:1609.09106}, 2016.

\bibitem{bertinetto2016learning}
Luca Bertinetto, Jo{\~a}o~F Henriques, Jack Valmadre, Philip Torr, and Andrea
  Vedaldi,
\newblock ``Learning feed-forward one-shot learners,''
\newblock in {\em NIPS}, 2016, pp. 523--531.

\bibitem{brock2018smash}
Andrew Brock, Theo Lim, J.M. Ritchie, and Nick Weston,
\newblock ``{SMASH}: One-shot model architecture search through
  hypernetworks,''
\newblock in {\em ICLR}, 2018.

\bibitem{zhang2018graph}
Chris Zhang, Mengye Ren, and Raquel Urtasun,
\newblock ``Graph hypernetworks for neural architecture search,''
\newblock in {\em International Conference on Learning Representations}, 2019.

\bibitem{nachmani2019}
Eliya Nachmani and Lior Wolf,
\newblock ``Hyper-graph-network decoders for block codes,''
\newblock in {\em Advances in Neural Information Processing Systems (NeurIPS)},
  2019.

\bibitem{nachmani2017learning}
Eliya Nachmani, Elad Marciano, Loren Lugosch, Warren~J Gross, David Burshtein,
  and Yair Be’ery,
\newblock ``Deep learning methods for improved decoding of linear codes,''
\newblock {\em IEEE Journal of Selected Topics in Signal Processing}, vol. 12,
  no. 1, pp. 119--131, 2018.

\bibitem{nachmani2020molecule}
Eliya Nachmani and Lior Wolf,
\newblock ``Molecule property prediction and classification with graph
  hypernetworks,''
\newblock {\em arXiv preprint arXiv:2002.00240}, 2020.

\bibitem{arikan2008channel}
Erdal Arikan,
\newblock ``Channel polarization: A method for constructing capacity-achieving
  codes,''
\newblock in {\em 2008 ISIT}. IEEE, 2008, pp. 1173--1177.

\bibitem{pamuk2011fpga}
Alptekin Pamuk,
\newblock ``An fpga implementation architecture for decoding of polar codes,''
\newblock in {\em ISWCS}. IEEE, 2011.

\end{thebibliography}

\end{document}